\documentclass[prr,twocolumn,amsfonts,amssymb,amsmath,floatfix,floats,a4paper,superscriptaddress]{revtex4-1}

\usepackage[margin=0.75in]{geometry}
\usepackage{graphicx}
\usepackage{amsmath}
\newcommand{\overbar}[1]{\mkern 1.5mu\overline{\mkern-3mu#1\mkern-1.0mu}\mkern 1.5mu}
\usepackage{scalerel}
\usepackage{appendix}

\begin{document}

\title{Footprints of quantum pigeons}

\author{Gregory Reznik}
\author{Shrobona Bagchi}
\affiliation{Raymond and Beverly Sackler School of Physics and Astronomy, Tel-Aviv University, Tel-Aviv 69978, Israel}
\author{Justin Dressel}
\affiliation{Institute for Quantum Studies, Chapman University, Orange CA 92866, USA}
\affiliation{Schmid College of Science and Technology, Chapman University, Orange CA 92866, USA}
\author{Lev Vaidman}
\affiliation{Raymond and Beverly Sackler School of Physics and Astronomy, Tel-Aviv University, Tel-Aviv 69978, Israel}
\affiliation{Institute for Quantum Studies, Chapman University, Orange CA 92866, USA}

\date{\today}

\begin{abstract}
We show that in the mathematical framework of the quantum theory the classical pigeonhole principle can be violated more directly than previously suggested, i.e., in a setting closer to the traditional statement of the principle. We describe how the counterfactual reasoning of the paradox may be operationally grounded in the analysis of the tiny footprints left in the environment by the pigeons. After identifying the drawbacks of recent experiments of the quantum pigeonhole effect, we argue that a definitive experimental violation of the pigeonhole principle is still needed and propose such an implementation using modern quantum computing hardware: a superconducting circuit with transmon qubits.
\end{abstract}


\maketitle

\section{Introduction}
Quantum paradoxes describe phenomena that would be impossible if Nature strictly obeyed classical physics. Quantum mechanics presents many paradoxes. A particular class of quantum paradoxes arises when we consider quantum systems between an initial preparation and final measurement. Notable examples of such pre- and postselection paradoxes include the three-box paradox \cite{Aharonov1991} where it is inferred that a particle with certainty  has been in two distinct locations simultaneously, and the Hardy paradox \cite{Hardy} where it is inferred that each particle of a particle-antiparticle pair has traveled through the same region of space without appearing there together.
A more recent example is the quantum pigeonhole paradox \cite{peculiar2013,aharonov2016} where one places a number of particles into a smaller number of boxes and infers that no two particles had occupied the same box. This latter paradox has prompted extensive discussion and several experimental implementations \cite{comment,reply,NMR,neutron,china2019}. We revisit this pigeonhole paradox and propose a conceptually stronger variation.  We also suggest that the existing experimental implementations have not yet definitively demonstrated the paradox.

The classical pigeonhole principle states that if one puts $N$ pigeons into $M$ pigeonholes, such that ${N>M}$, then there must be at least one pigeonhole that contains more than one pigeon. It was formulated by Dirichlet in the 19th century \cite{diri} and is widely used in number theory and combinatorics. The principle seems obvious and formalizes the fundamental concept of counting, yet it can be apparently violated by pre- and postselected quantum systems.

\section{Elements of reality}
To demonstrate a quantum violation of the classical pigeonhole principle one prepares a particular superposition of $N$ (quantum) pigeons distributed into $M$ (classical) holes, then later measures another particular superposition of the $N$ pigeons. In between the preparation and a successful postselection one then predicts with certainty that any particular hole does not contain more than one pigeon. Moreover, this surprising prediction may be checked experimentally by placing a probe to count the pigeons in any box. A somewhat weaker failure of the classical pigeonhole principle can be obtained when the holes are also quantum (e.g. spin states), since it is less surprising that intrinsically quantum features do not follow classical rules.

Even if the ``holes'' in such a scenario are classical, we still have to clarify the meaning of a ``quantum pigeon being in a hole''. Standard quantum mechanics does not have a clear answer to the question: Where was a particle in between a preselection and postselection?
 In classical physics, the statement ``this pigeon is in that hole'' can be tested in parallel by many different measurements that do not affect the situation. We do not assume this for quantum pigeons, because measurements performed on a quantum object generally change its state. A quantum pigeon can  be prepared in a superposition of several spatial locations, which also makes statements about such a pigeon occupying a particular hole not clearly defined. The exception is when a quantum pigeon is described by a well-localized wave packet with support only in one hole, in which case no paradoxical behavior arises. So, we need to carefully define what we mean by a quantum pigeon occupying a particular hole. We will use the following definition \cite{Elements_of_reality}:
\begin{quote}
    If we can infer with certainty [that] the result of a measurement at time $t$ of an observable $C$ equals to $c$, then $C=c$ is an element of reality.
\end{quote}
In our case: If we can infer with certainty that the measurement at time $t$ of the presence of the pigeon in a particular hole would yield a positive result, then the pigeon was in the hole at time $t$.

For a quantum system that is only preselected, a measurement outcome $C=c$ will be obtained with certainty only if the system is prepared in an eigenstate of $C$. However,  when the system is both pre- and postselected, the condition for obtaining measurement outcome $C=c$ with certainty is different.

For a system preselected in a state  $|\Psi\rangle$ and postselected in a state $|\Phi\rangle$, the probability for a particular result of an intermediate measurement is given by the Aharonov-Bergmann-Lebowitz (ABL) formula \cite{ABL}
\begin{align}\label{ABL}
    {\rm Prob}(C=c) ~  =~\frac{| \langle \Phi | {\rm \bf P}_{C = c} |\Psi\rangle|^2}{| \langle \Phi | {\rm \bf P}_{C = c} |\Psi\rangle|^2+| \langle \Phi | {\rm \bf P}_{C \neq c} |\Psi\rangle|^2}~.
\end{align}
Thus, the requirement for $C=c$ to be an element of reality, i.e. ${\rm Prob}(C=c) = 1$, becomes:
\begin{align}\label{element_reality}
    \begin{matrix}
    C=c\ {\rm is \ an\  element}\\
   {\rm \ \ \ ~~  of\ reality}
    \end{matrix}
   ~~ \Longleftrightarrow
   ~ \begin{cases}
    \langle \Phi | {\rm \bf P}_{C = c} |\Psi\rangle \neq 0\\
	\langle \Phi | {\rm \bf P}_{C \neq c} |\Psi\rangle = 0
    \end{cases}.
\end{align}
Provided that the postselection becomes impossible when $C\neq c$, we can infer that $C$ would be measured to be $c$ with certainty when the postselection succeeds.

The measurement in this definition is understood as counterfactual, i.e., it did not necessarily happen. However, it is assumed that if the measurement of $C$ had been performed, then it must have been the only measurement on the system between the pre- and postselection. Making more than one measurement would change the scenario and disrupt the inference. Even if the parts of the system are far away, performing measurement on one part can influence possible outcomes of the  measurement of other parts.
\section{Violating the pigeonhole principle with $N$ pigeons in two holes}

The classical pigeonhole principle is a global statement about all holes: there should exist at least one hole with a particular property (having more than one pigeon). We will now show for quantum mechanics that given a particular pre- and postselection scenario we can infer with certainty that we will not find more than one pigeon in a single hole that we check. The paradoxical situation is that we are certain not to find more than one pigeon in {\it any} one of the holes we try, no matter how many times we try to find a hole containing more than one pigeon. Nature seems to conspire against the experimenter by always hiding multiple pigeons from view, provided that the experimenter only checks one box at a time and obtains a successful postselection.

We consider $N$ pigeons placed in two pigeonholes $A$ and $B$. The pigeons may be partitioned into subsets of labeled pairs $\{j,k\}$, triples $\{j,k,l\}$ and so forth.
The statement that pigeonhole $X$ contains more than one pigeon then corresponds to the projection operator
\begin{multline}\label{>1}
{\rm \bf P}^{\scaleto{>1}{4pt}}_X =
\sum_{\{j,k\}}
\prod_{m = j,k} {\rm \bf P}^{(m)}_{X} \prod_{m \neq j,k} {\rm \bf P}^{(m)}_{\overbar{X}}+ \\
\sum_{\{j,k,l\}}
\prod_{m = j,k,l} {\rm \bf P}^{(m)}_{X} \prod_{m \neq j,k,l} {\rm \bf P}^{(m)}_{\overbar{X}}+
...+
\prod_{m} {\rm \bf P}^{(m)}_X
,
\end{multline}
where ${\rm \bf P}^{(j)}_X = | X \rangle _j\langle X |_j$ denotes the projection on a state in which pigeon $j$ is present in hole $X$, ${\rm \bf P}^{(j)}_{\overbar{X}} = {\rm \bf I}^{(j)} - {\rm \bf P}^{(j)}_X$ denotes the complementary projection on the state in which pigeon $j$ is not present in hole $X$, and the summations are over the possible subsets of two or more pigeons. The negation of ${\rm \bf P}^{\scaleto{>1}{4pt}}_X$ is that pigeonhole $X$ does not contain more than one pigeon, ${\rm \bf P}^{{\scaleto{\leq 1}{4pt}} }_X = {\rm \bf I} - {\rm \bf P}^{\scaleto{>1}{4pt}}_X$, i.e., pigeonhole $X$ contains either one or zero pigeons:
\begin{align}
{\rm \bf P}^{\scaleto{\leq 1}{4pt}}_X =
\prod_{m} {\rm \bf P}^{(m)}_{\overbar{X}}+
\sum_{\{j\}}
{\rm \bf P}^{(j)}_X  \prod_{m \neq j} {\rm \bf P}^{(m)}_{\overbar{X}}
.
\end{align}

For the pigeonhole principle to fail, the observable $C = {\rm \bf P}^{\scaleto{\leq 1}{4pt}}_X$ should be inferred to have the value $c=1$ with certainty for either choice of $X$. Since ${\rm \bf P}_{C = 1} = {\rm \bf P}^{\scaleto{\leq 1}{4pt}}_X$ and ${\rm \bf P}_{C \neq 1} = {\rm \bf P}^{\scaleto{>1}{4pt} }_X$, (\ref{ABL}) produces the following requirements:
\begin{align}\label{less_1}
\langle\Phi | {\rm \bf P}^{\scaleto{\leq 1}{4pt} }_X | \Psi\rangle \neq 0, ~~\langle\Phi | {\rm \bf P}^{\scaleto{>1}{4pt} }_X | \Psi\rangle = 0.
\end{align}

\subsection{How to place four pigeons in two holes with not more than one pigeon in each hole}

We now demonstrate the failure of the pigeonhole principle with four pigeons in two pigeonholes, modeled as four particles in two boxes. (We do not expect to perform experiments with real quantum pigeons.) A single measurement of the presence of more than one particle in any of the holes, yields with certainty ${\rm \bf P}^{\scaleto{>1}{4pt}}_X=0$. We prepare the particles in the initial state
\begin{align}\label{eq:1}\nonumber
|\Psi \rangle=\frac{1}{\sqrt{3}}\Big(
| A \rangle _1| A \rangle _2| A \rangle _3| A \rangle _4 +
| A \rangle _1| A \rangle _2| B \rangle _3| B \rangle _4+\\
| B \rangle _1| B \rangle _2| B \rangle _3| B \rangle _4
\Big),
\end{align}
then  postselect the particles in the final state:
\begin{align}\label{eq:2}\nonumber
|\Phi \rangle=\frac{1}{\sqrt{3}}\Big(
| A \rangle _1| A \rangle _2| A \rangle _3| A \rangle _4 -
| A \rangle _1| A \rangle _2| B \rangle _3| B \rangle _4+\\
| B \rangle _1| B \rangle _2| B \rangle _3| B \rangle _4
\Big).
\end{align}
Our requirements for (\ref{element_reality}) are then satisfied:
\begin{align}\label{calc_1}
\langle\Phi| {\rm \bf P}^{\scaleto{>1}{4pt}}_A|\Psi\rangle= \frac{1}{3}(1-1)=0,
~~
\langle\Phi| {\rm \bf P}^{\scaleto{\leq 1}{4pt}}_A|\Psi\rangle= \frac{1}{3}.
\end{align}
Similarly,
\begin{align}
\langle\Phi| {\rm \bf P}^{\scaleto{>1}{4pt}}_B|\Psi\rangle=0,~~\langle\Phi| {\rm \bf P}^{\scaleto{\leq 1}{4pt}}_B|\Psi\rangle = \frac{1}{3}.
\end{align}
If we were to try to find more than one particle in box $A$ between pre- and postselection, then we would be certain to fail. Similarly, if we were to try to find more than one particle in box $B$, we would be certain to fail. No matter how many times we attempt to find multiple particles in any single box, we would fail.

In fact, our example demonstrates even stronger violation of classical reasoning. We put four particles in two boxes such that there are no  particles at all in every box! That is, an observable $C' = {\rm \bf P}^{\scaleto{=0}{4pt}}_X$ testing whether there are zero particles in box $X$ will show with certainty that there are none, $c'=1$. Indeed, the complement ${\rm \bf P}^{\scaleto{>0}{4pt}}_X = {\rm \bf I} - {\rm \bf P}^{\scaleto{=0}{4pt}}_X$ corresponding to $c' \neq 1$ has the form
\begin{align} \label{>0}
{\rm \bf P}^{\scaleto{>0}{4pt}}_X=
\sum_{\{j\}}
{\rm \bf P}^{(j)}_X  \prod_{m \neq j} {\rm \bf P}^{(m)}_{\overbar{X}}+ {\rm \bf P}^{\scaleto{>1}{4pt}}_X,
\end{align}
and we obtain our requirements for (\ref{element_reality})
\begin{align}
\langle\Phi| {\rm \bf P}^{\scaleto{>0}{4pt}}_X|\Psi\rangle=0,~~
\langle\Phi| {\rm \bf P}^{\scaleto{=0}{4pt}}_X|\Psi\rangle
= \frac{1}{3}.
\end{align}
Note that these results strongly depend on the exact definition of the measurements (\ref{>1}) and (\ref{>0}). If we ask a different question, ``Are there exactly four particles in box $X$?'', then the outcome will be yes with certainty, ${\rm \bf P}^{\scaleto{=4}{4pt}}_X=1$ for both boxes $X$.

\subsection{How to place $N$ pigeons in two holes with not more than $K$ pigeons in a hole}

Let us consider how to generalize this result. We discussed cases with no particles in a box and with no more than one particle in a box.  Classically, it is possible to distribute $N$ particles between two boxes with no more than $K$ particles in each box only if $N \leq 2K$. We find that in quantum mechanics it is also possible when $N > 2K$, except for one special case in which $N=2K+1$. Indeed, When $N>2K+1$ we can use the same method.  We prepare the particles in the initial state
\begin{align}\label{eq:1+}
|\Psi \rangle=\frac{1}{\sqrt{3}}\Big(
\prod_{n=1}^N
| A \rangle_n +\prod_{n=1}^{K+1}
| A \rangle_n\prod_{m=K+2}^{N}
| B \rangle_m
+
\prod_{n=1}^{N}
| B \rangle_n
\Big),
\end{align}
then  postselect the particles in the final state:
\begin{align}\label{eq:2+}
|\Phi \rangle=\frac{1}{\sqrt{3}}\Big(
\prod_{n=1}^N
| A \rangle_n -\prod_{n=1}^{K+1}
| A \rangle_n\prod_{m=K+2}^{N}
| B \rangle_m
+
\prod_{n=1}^{N}
| B \rangle_n
\Big).
\end{align}
In case $N=2K+1$ this method does not work for box $B$ and straightforward calculation shows that no successful method exists. For arbitrary pre and postselection both conditions of ${\rm \bf P}^{\scaleto{>K}{4pt} }_A=0$ and ${\rm \bf P}^{\scaleto{>K}{4pt}}_B=0$ can only be satisfied if the whole preselected state is orthogonal to the postselected state, which is impossible. Therefore, there is no example of placing three particles in two boxes such that no box contains more than one particle.

Note, that there is no limitation when number of boxes $M>2$. If $N\leq KM$ then there is even a classical solution for putting particles such that not more than $K$ particles are present in any box. When $N>KM$, then the quantum solution is the preselection of state (\ref{eq:1+}) and postselection of state (\ref{eq:2+}). The only exception is $N=1$ and $K=0$.

\subsection{How to place indistinguishable pigeons in two holes with not more than one pigeon in each hole}

The failure of the pigeonhole principle can be demonstrated also for quantum indistinguishable particles. In case of identical particles, using a Fock state representation is more convenient. For example, the projection in (\ref{>1}) becomes
\begin{align}
{\rm \bf P}_X^{\scaleto{>1}{4pt} }=
| 2 \rangle_X\langle 2|_X+
| 3 \rangle_X\langle 3 |_X+
...+| N  \rangle_X\langle  N  |_X,
\end{align}
where $|n\rangle_X$ denotes the Fock state with $n$ identical particles in the box $X$.
Similarly, the pre- and postselection states in (\ref{eq:1}) and (\ref{eq:2}) become
\begin{align}
|\Psi \rangle=\frac{1}{\sqrt{3}}\Big(
|4\rangle_A |0 \rangle_B  +|2\rangle_A |2 \rangle_B +|0\rangle_A |4 \rangle_B \Big),
\end{align}
and
\begin{align}
|\Phi \rangle=\frac{1}{\sqrt{3}}\Big(
|4\rangle_A |0 \rangle_B  -|2\rangle_A |2 \rangle_B +|0\rangle_A |4 \rangle_B \Big),
\end{align}
and lead to the same situation. The measurement of the presence of more than one pigeon in any hole $X$ yields ${\rm \bf P}^{\scaleto{>1}{4pt}}_X=0$ with certainty. Moreover, the measurement of more than zero pigeons in each hole also yields ${\rm \bf P}^{\scaleto{>0}{4pt}}_X=0$ with certainty. And it can be shown in the same way that the generalization of  section III.B for $K$ pigeons in the hole  and more than two holes hold for indistinguishable pigeons too.

\section{Previous proposals to violate the pigeonhole principle}

We have presented a method of violating the pigeonhole principle with quantum pre- and postselected particles. Our example logically fits the classical pigeonhole principle definition more directly than previous proposals \cite{peculiar2013,aharonov2016}. However, our proposal has a serious weakness for experimental verification. As mentioned above, the meaning of an observable $C$ that asks whether there is more than one pigeon in a particular hole is that there is a measuring device capable of displaying only one of two readings: `yes', there is more than one pigeon, or `no', there is no more than one pigeon. That is, the quantum measurement should not provide $K$, the exact number of pigeons in the hole, but instead only two readings: $K> 1$ and $K\leq 1$.

The physical implementation of such a measurement requires that the measuring device must be affected exactly in the same way when we have two pigeons in the hole and when we have three or four pigeons in the hole. Similarly, it must be affected exactly in the same way for either one or zero pigeons. 
While it is not unthinkable to arrange an effective interaction that achieves a similar response for two or more quantum pigeons, most basic physical interactions are bi-particle couplings so it is challenging to ensure the needed insensitivity to particle number. Thus, the previous proposals for demonstrating the failure of the pigeonhole principle (which are based on bi-particle interactions) are still attractive from an experimental point of view even if their definitions do not fit the exact wording of the classical pigeonhole principle.

\subsection{How to place $N$ pigeons in two holes such that no hole contains two pigeons}
The pigeonhole principle tells us that after placing $N>2$ pigeons in two holes there should be at least one hole with more than one pigeon. More than one is at least two, so a slightly weaker test is to check whether there is at least one hole with two pigeons.
Classically, there is no difference, since one can always find two pigeons as a subset of more than two, so it is sufficient to show that no holes have two pigeons to demonstrate a violation of the pigeonhole principle. However, in quantum mechanics there can be a difference between asking for exactly two pigeons and asking for two or more.

In \cite{peculiar2013}
a situation  in which $N>2$ particles are placed into two boxes such that no box contains a pair of particles was presented. Since this situation should not occur classically, this weaker test still implies a failure of the pigeonhole principle. To achieve this the following states were pre- and postselected:
\begin{multline}\label{pre13}
|\Psi \rangle=\frac{1}{\sqrt {2^{N+1}}}\bigg [
\prod_{n=1}^N(
| A \rangle_n - i| B \rangle_n) +
\prod_{n=1}^N(
| B \rangle_n - i| A \rangle_n
)\bigg ],
\end{multline}
 \begin{align}\label{post13}
|\Phi \rangle=\frac{1}{\sqrt {2^{N}}}
\prod_{n=1}^N(
| A \rangle_n +| B \rangle_n
).
\end{align}
In this situation we can claim that every pair of particles $\{j,k\}$ is not present together in any particular box $X$.  For every pair $\{j,k\}$, the probability to find the pair in  any box $X$  vanishes. Indeed, we obtain:
\begin{align}\label{ijX1}
\langle \Phi|{\rm \bf P}^{\{j,k\}}_X |\Psi \rangle=\frac{(1-i)^{N-2}}{2^{N-\frac{1}{2}}}(1^2+(-i)^2)=0,
\end{align}
\begin{align}\label{ijX2}
\langle \Phi|{\rm \bf I} - {\rm \bf P}^{\{j,k\}}_X |\Psi \rangle=\frac{-i(1-i)^{N-2}}{2^{N-\frac{3}{2}}}\neq 0,
\end{align}
where ${\rm \bf P}^{\{j,k\}}_X={\rm \bf P}^{(j)}_X {\rm \bf P}^{(k)}_X$.

While this test is classically equivalent to testing that there is certainly no more than one particle in each box, this is not true quantum mechanically. Indeed, if we perform a similar test for the presence of exactly three particles in a particular box using the same pre- and postselections, then we have nonvanishing probability to find them. The ABL formula (\ref{ABL}) yields
\begin{multline}\label{ijkX1}
 {\rm Prob} ({\rm \bf P}^{\{j,k,l\}}_X =1)=
\frac{| \frac{(1-i)^{N-4}}{2^{N-\frac{1}{2}}} |^2}
{|\frac{(1-i)^{N-4}}{2^{N-\frac{1}{2}}}|^2+
|-5\frac{(1-i)^{N-4}}{2^{N-\frac{1}{2}}}  |^2}\\
=\frac{1}{26}
.
\end{multline}

This is why this example is formally not as strong as our first example, even though the classical pigeonhole principle is violated in both.
Nevertheless, this example has an intriguing physical meaning in quantum mechanics. The implied phenomenon is that if particles $j$ and $k$ would normally interact with each other when both present in box $X$, then in the specified pre- and postselected situation the particles would apparently not interact.  Moreover,  provided the particle interactions are weak enough, the particular pre- and postselection effectively switches off {\it all} bi-particle interactions while preserving interactions between larger numbers of particles.

To explain this, recall a theorem connecting strong and weak measurements \cite{Aharonov1991}. If the result of a strong measurement of some variable obtains a particular eigenvalue with certainty, then the weak value is equal to this eigenvalue. Thus, for all pairs of particles and for both boxes $\left ({\rm \bf P}^{\{j,k\}}_X\right )_w =0$ holds.
Weak values characterize effective weak coupling and since weak coupling does not disturb significantly the two-state vector description of the pre- and postselected particles, these null weak values remain small even when all (weak) couplings are present.
This is arguably the most interesting physical implication of the quantum pigeonhole effect.


\subsection{How to violate the pigeonhole principle without entanglement}
In \cite{aharonov2016}  another proposal for the failure of the pigeonhole principle (which attracted significantly more attention) was presented. This variation showed that the failure of the pigeonhole principle can occur even in systems without entangled pre- and postselections (see also \cite{comment,reply}). The lack of entanglement makes this variation particularly attractive for experimental implementation. Consider the following pre- and postselected states:
\begin{align}\label{pre16}
|\Psi \rangle=
\frac{1}{\sqrt {2^{N}}}
\prod_{n=1}^N(
| A \rangle_n +| B \rangle_n
),
\end{align}
 \begin{align}\label{post16}
|\Phi \rangle=
\frac{1}{\sqrt {2^{N}}}
\prod_{n=1}^N(
| A \rangle_n +i| B \rangle_n
).
\end{align}
The pre- and postselected states are completely separable; nevertheless, the probability to find any particular pair of particles in the {\it same} box is zero.

As before, this statement is correct only when one pair is tested. Moreover, unlike the previous example it is correct only if the boxes A and B are not distinguished.  The projection operator corresponding to this measurement is
\begin{align}
{\rm \bf P}^{\{j,k\}}_{\rm same} =
{\rm \bf P}_A^{\{j,k\}} +
{\rm \bf P}_B^{\{j,k\}}.
\end{align}
It tells us whether or not  the   particles $j$ and $k$ are present in the same box without providing information about which box they are in.

For our pre- and postselected states we obtain for every pair ${j,k}$:
\begin{align}
\langle\Phi | {\rm \bf P}^{\{j,k\}}_{\rm same} | \Psi\rangle =\frac{(1-i)^{N-2}}{2^N}(1^2 +i^2)=0,
\end{align}
\begin{align}
\langle\Phi | {\rm \bf I} - {\rm \bf P}^{\{j,k\}}_{\rm same} | \Psi\rangle =\frac{-i(1-i)^{N-2}}{2^{N-1}} \neq 0.
\end{align}

Similarly to the previous example, and unlike  classical physics, even if we are sure not to find any pair in the same box, we might still find three particular particles being in the same (without knowing which) box. The ABL formula (\ref{ABL}) for such a case yields:

\begin{multline}\label{prob3}
 {\rm Prob} ({\rm \bf P}^{\{j,k,l\}}_{\rm same} =1)=
\frac{
|\frac{(1-i)^{N-4}}{2^{N-1}} |^2}
{|\frac{(1-i)^{N-4}}{2^{N-1}} |^2+
|-3\frac{(1-i)^{N-4}}{2^{N-1}}   |^2}\\
=\frac{1}{10}
.
\end{multline}

\section{Experiments demonstrating the pigeonhole paradox}

Testing for the presence of particles is challenging,
so the most promising experimental implementation for violating the pigeonhole principle is that of example \cite{aharonov2016}, since it tests the  particle pair interactions rather than the locations of the particles. Moreover, this implementation has pre- and postselected  separable states that are more easily arranged.
Still, the experiment is very difficult, since the natural coupling between pairs of particles is very weak.

There are now several experimental papers that claim to demonstrate the violation of the pigeonhole principle. In \cite{NMR} ``NMR investigation of pigeonhole effect'' quantum gates that schematically simulate the pigeonhole experiment were implemented. Quantum simulation, i.e. performing a sequence of quantum gates that formally model the pigeonhole experiment, is not a compelling demonstration. In NMR experiment there is no direct connection between logical qubit and physical local system.
More physical implementations were performed with neutrons \cite{neutron},  and,  more recently, with photons \cite{china2019}. We argue that all these experiments are not yet satisfactory for definitively demonstrating the quantum pigeonhole effect.

A direct demonstration of \cite{aharonov2016} can be generally divided into the following tasks. 

i) Prepare $N$ particles in the prescribed state (\ref{pre16}) and then postselect the particles in the state (\ref{post16}).

ii) Add a strong interaction between randomly chosen pair of particles conditioned on their presence in the same box. Upon pre- and postselection according to (i), show that this interaction is suppressed.

iii) An alternative to (ii) that is closer to the spirit of the original classical pigeonhole principle is to strongly measure, using external devices, that a randomly chosen pair does not share the same box (without distinguishing the boxes).

iv) Replace a strong interaction as in (ii) by    a weak bi-particle interaction, but make it  between all pairs of  particles. Show that upon pre- and postselection as in (i), the effect of the interactions almost disappears (becomes second-order in the weak disturbance).

Task (i) for \cite{aharonov2016} is simple and there is no doubt that it was demonstrated, even if it was not specifically reported in the experimental papers on the quantum pigeonhole effect. However, it is clear that (i) by itself is not sufficient. From a physics point of view, task (iv) might be the most interesting experiment; however, we have not seen a convincing implementation of it (despite some claims made in \cite{china2019}). Performing task (ii) or task (iii) is the most important to be able to claim that the pigeonhole effect was demonstrated. We will now analyze to which extent they were achieved.

A common weakness of the existing pigeonhole experiments is that the ``holes'' are usually spin or polarization states. These degrees of freedom are manifestly quantum concepts, so it is not so strange that they fail to fulfill a classical principle. Nevertheless, these demonstrations do show the conceptual failure of the principle. Spin can be up or down. If we have more than two particles, classical counting logic still tells us that there should be at least one pair of particles with the same spin state.

\subsection{Demonstration of the failure of the pigeonhole principle with neutrons}

Let us first discuss experiment \cite{neutron}, which uses the $z$ component of a neutron's spin to encode which ``box'' it occupies, with $\sigma_z=1$ signifying hole $A$ and $\sigma_z=-1$ signifying hole $B$. The experiment includes a source of individual neutrons and devices that prepare and postselect the required spin-polarization states, so task (i) is achieved.

No direct demonstration of task (ii) for pairs of neutrons was performed. Instead, a careful measurement of the weak value of $\sigma_z$ was performed for each neutron.
The following argument that this weak measurement is  sufficient for demonstrating the failure  of a pigeonhole principle was provided:

a) The weak measurement provided the weak value of the spin component of the pre- and postselcted neutrons, $(\sigma_z)_w=i$.

b) For a product of variables related to separable particles (non-entangled pre and postselection states), the weak value of a product is a product of weak values: $(O^{(j)}O^{(k)})_w=(O^{(j)})_w(O^{(k)})_w$. Thus,
\begin{align}\label{i^2}
(\sigma_z^{(j)}\sigma_z^{(k)})_w=(\sigma_z^{(j)})_w(\sigma_z^{(k)})_w=i^2=-1.
\end{align}
c) For dichotomic variables, if a weak value is equal to an eigenvalue, then this eigenvalue, if measured, will be obtained with certainty, i.e. it is an element of reality. (This theorem appeared first in \cite{Aharonov1991}.)

Therefore, for every pair of particles $j$ and $k$ we have an element of reality  $\sigma_z^{(j)}\sigma_z^{(k)}=-1$. The interpretation of this is that these two particles have opposite spin $z$ components, which corresponds to particles ``being in different holes''.  This inference thus achieves task (iii).

The difficulty with this experiment is that the joint spin measurement was not actually performed since no two neutrons ever coexisted in the measurement apparatus. Only single-particle weak measurements of $(\sigma_z)_w=i$ were performed. The result $\sigma_z^{(j)}\sigma_z^{(k)}=-1$ was only {\it inferred}, which was only possible because the pre- and postselections were known to be separable. 
A more convincing demonstration should be performed without relying on any prior information about the pre- and postselected states of the particles. Without this information, Eq.~(\ref{i^2}) of step b) does not hold, so the measured weak values do not directly provide a demonstration of the failure of the pigeonhole principle.

Indeed, the information that  the pre- and postselected states are not entangled is crucial. Consider the following pre- and postselected states of $N$ particles:
\begin{align}\label{pre-ent}
|\Psi \rangle=\frac{1}{\sqrt 2}\left (\prod_{n=1}^N
|{\uparrow} \rangle_n  + \prod_{n=1}^N|{\downarrow}\rangle_n  \right )    , \end{align}
 \begin{align}\label{post-ent}
|\Phi \rangle=\frac{1}{\sqrt 2}\left (\prod_{n=1}^N
|{\uparrow} \rangle_n  + i \prod_{n=1}^N|{\downarrow}\rangle_n  \right ).
\end{align}
For this pre- and postselected state of $N$ particles for every particle $n$ we have $(\sigma_z)^{(n)}_w=i$, but for every pair we have  $ (\sigma_z^{(j)}\sigma_z^{(k)}  )_w=1$.
This example, in which there is no failure of the pigeonhole principle despite every particle having value $(\sigma_z)_w=i$, shows that the experiment \cite{neutron} does not provide an unconditional demonstration of the pigeonhole principle.

\subsection{Demonstration of the failure of the pigeonhole principle with photons}

Let us now turn to experiment \cite{china2019}. From the abstract one can understand that it achieves almost everything, including weak measurements, i.e., task (iv). However, in the summary it was admitted that there was no direct demonstration: ``We implement the desired measurement indirectly by analyzing
the measurement effects order by order and reveal the paradox will not survive under high-order measurement.'' It is not clear what it might mean ``order by order'' since only strong couplings were described in this experiment. As such, we see mainly an attempt to perform task (ii), but we feel that even this was not done in a fully  satisfactory way. We see the following weaknesses of the experiment.

a) {\it Instead of showing that properly pre- and postselected particles have the property of ``not being in the same hole'', it was shown that particles with the desired property were never found with proper postselection.}

In task (iii) we are supposed to randomly choose two properly pre- and postselected particles and perform a strong measurement to show that $\sigma_z^{(j)}\sigma_z^{(k)}=-1$. Instead, the experiment was done in such a way that the particles fulfilling $\sigma_z^{(j)}\sigma_z^{(k)}=1$ had zero probability of being properly postselected. Indeed, this is what the experiment showed: none of the properly pre- and postselected particles with this property were found in the experiment. This finding supports the claim that properly postselected particles do not fulfill $\sigma_z^{(j)}\sigma_z^{(k)}=1$, but does not {\it demonstrate} the paradoxical situation of the failure of the pigeonhole principle. There were no observed and detected photons pre and postselected in the states   corresponding to the failure of the pigeonhole principle.

Note that adding number-resolving photon detectors to allow observing the properly postselected photons that both arrive at the same port (corresponding to different polarizations) could solve the problem. In this case, however, the detection would not just tell us that the photons do, or do not have the same polarization, as was proposed in \cite{aharonov2016}. Instead, the detector clicks will not reveal which polarization they are if they are the same, but they would reveal the polarizations when they are different. This difference, however, is not crucial, since it does not remove the paradoxical feature of the original proposal.

b) {\it The paradox is defined for  distinguishable particles, but the experiment was an interference of identical particles.}

The pigeonhole principle failure in \cite{aharonov2016} is defined for distinguishable particles. For any particular pair $(j,k)$ we know that $\sigma_z^{(j)}\sigma_z^{(k)}=-1$. If the particles are identical and cannot be distinguished, then there is no meaning for this statement. Instead of coupling between distinct particles, the core mechanism of the measurement in experiment \cite{china2019} is Hong-Ou-Mandel interference, which requires identical bosons. The example  \cite{aharonov2016} can be designed with identical particles if they are  distinguishable by some degrees of freedom, such as a mode in the interferometer which identifies them. Indeed, experiment \cite{china2019} starts with three identical photons, but in different modes which could identify the particles of the pigeonhole paradox. The problem is that their measuring device, the polarization beam splitter, mixes the modes, i.e. scrambles the identity. How can we demonstrate that particular particles do not have the same polarization when the measurement loses the identity of the particles?

c) {\it The demonstrated lack of disappearance of the effect of the interactions in the pigeonhole setup for strong measurements of two pairs of particles is not relevant, since it is expected only for weak coupling.}

The experiment \cite{china2019} was also performed with coupling (interference) of three particles, testing the presence of all three particles sharing the same spin state. However, in \cite{aharonov2016} it was never claimed that there are no three pigeons in the same hole, the claim was only about pairs (see also discussion in \cite{comment,reply}).

This fact is clearly demonstrated by (\ref{prob3}). The probability of finding three particles in the same box is nonzero and is equal to $\frac{1}{10}$. The experiment indeed demonstrated that there are sets of three photons with the same polarization, confirming (\ref{prob3}), but this is a demonstration of a situation which is not paradoxical.

\section{Implementation proposals}

In the previous section we provided more criticism of existing experiments than solutions. A convincing demonstration of the pigeonhole principle is a very difficult task. A proper implementation of the example \cite{aharonov2016} requires the measurement of nonlocal variables like parity \cite{parity}. For example, the setup of the recent implementation \cite{PRL-nl} of nonlocal measurement \cite{AAV86} could allow a more satisfactory experiment showing the failure of the pigeonhole principle in the version with ``holes'' still being spin states. Indeed, in \cite{PRL-nl} there was the experimental realization of a pre- and postselected pair of particles fulfilling a requirement similar to the type we are looking for: $\sigma_x^{(j)}\sigma_z^{(k)}=-1$. Importantly, this property was measured, so this setup could be adapted to demonstrate task (iii) of the quantum pigeonhole effect.

Modern superconducting quantum computation circuits could  enable a demonstration of not only task (iii) for \cite{aharonov2016}, but also task (iv).
One would encode the ``boxes'' as distinct energy states of an anharmonic oscillator, such as a transmon \cite{koch2007charge}.   Since these distinct states correspond to mesoscopic collective charge oscillations along superconducting wires at the micron scale, their oscillation energies are somewhat more defensible classical boxes than the intrinsic spin states of individual quantum particles. Different oscillators need not share identical energies, but their lowest two energy states can be arranged to fall within the same energy intervals, and thus ``share the same energy boxes'' in a way analogous to how two classical particles with slightly different positions can share the same spatial box.

Direct parity measurements are possible for pairs of transmons and are already being experimentally implemented for the purpose of quantum error correction \cite{QuantumErrorTracking2019}, where the parity measurements are used to stabilize entangled code spaces and track bit errors. Thus, one could directly measure the needed negative parities, corresponding to $\sigma_z^{(j)}\sigma_z^{(k)} = -1$, for randomly chosen pairs of pre- and postselected transmons. This experiment will have a conceptual advantage relative to \cite{PRL-nl} since here the pointer is an external measuring device, and not another degree of the measured photons.

\begin{figure}
    \centering
    \includegraphics[width=\columnwidth,height=0.39\textwidth]{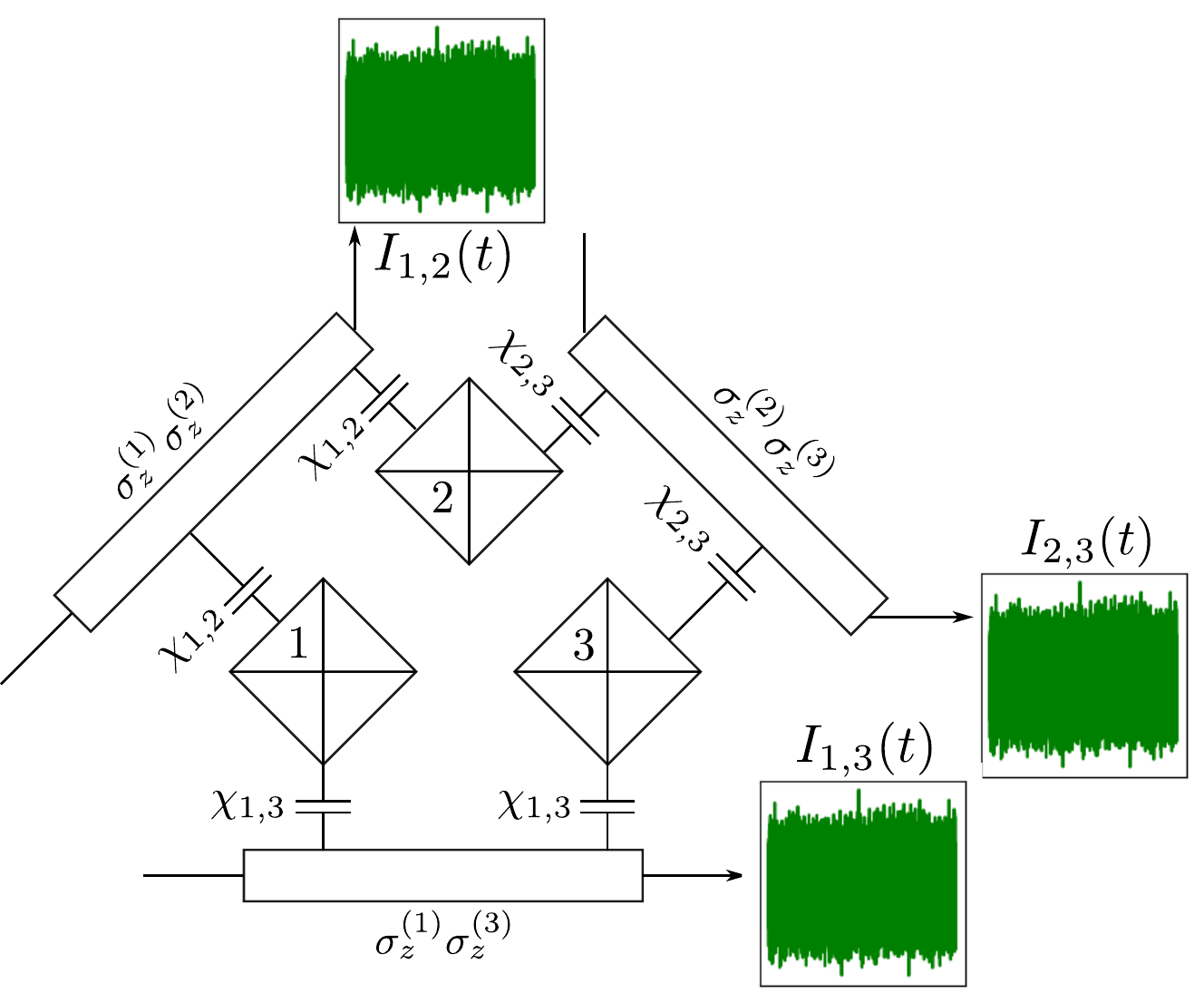}
    \caption{Proposed setup for demonstrating the quantum pigeonhole paradox with superconducting transmons. Each pair $(j,k)$ of three transmons $j,k=1,2,3$ is coupled dispersively to parity readout resonators. The transmon states determine the dispersive shifts $\chi_{j,k}$ of the resonator frequencies. Pumping the readout resonators will produce signals $I_{j,k}(t)$. For task (iii), one resonator is pumped for a long duration to projectively measure a parity eigenvalue $\sigma_z^{(j)}\sigma_z^{(k)} = -1$. For task (iv), all three resonators are pumped for a short time to measure the parity weak values of -1.}
    \label{fig:schematic}
\end{figure}

Figure~\ref{fig:schematic} illustrates how to implement the needed parity measurements for the quantum pigeonhole paradox with superconducting transmons. Each pair $(j,k)$ of three transmons $j,k=1,2,3$ is coupled dispersively to parity readout resonators. The resonator frequencies shift by amounts $\pm\chi_{j,k}$ conditioned on specific transmon energies. The shifts are tuned such that transmons coupled to the same resonator produce the same shifts for the same energies. Therefore, the resonances for the odd transmon subspace (01 and 10) will produce no net frequency shift $\chi_{j,k} - \chi_{j,k}  = 0$. In contrast, the even subspace (00 and 11) will produce distinct shifts of $\pm2\chi_{j,k}$ that are detuned beyond the resonator linewidth. Therefore, pumping the readout resonators on the common resonance for the odd subspace allows the two parity subspaces to be distinguished.
Pumping for a short duration produces an integrated signal $\bar{I}_{j,k}$ corresponding to a weak measurement of the parity $\sigma_z^{(j)}\sigma_z^{(k)}$, while pumping for a long duration produces a projective measurement.
To achieve task (iii) of \cite{aharonov2016}, one would select a random readout resonator and pump it for a long duration. We expect that for all successful postselections only the parity eigenvalue $\sigma_z^{(j)}\sigma_z^{(k)} = -1$ will be observed. This procedure can then be repeated for random pairs as many times as desired. Note that this procedure will work only if a single pair is measured at a time. If multiple resonators are pumped simultaneously, then no definite parity results will be obtained.

To achieve task (iv), i.e. to demonstrate that $\sigma_z^{(j)}\sigma_z^{(k)} = -1$ for all pairs simultaneously, one would instead pump all three resonators for a very short duration and record all signals. After averaging many successful postselections together, one would obtain the parity weak values $(\sigma_z^{(j)}\sigma_z^{(k)})_w= -1$ for all three pairs. The short pump duration prevents appreciable disturbance so that these weak values can be observed. As indicated before, a weak value (of our dichotomic variable) $(\sigma_z^{(j)}\sigma_z^{(k)})_w= -1$ implies that if this variable were strongly measured, then the measurement would report the same value with certainty.

An experiment of this type is on the cutting edge of current technology. Tuning the dispersive shifts of the six transmon-resonator couplings $\chi_{j,k}$ while allowing the individual qubit control to perform the needed pre- and postselections is a nontrivial engineering task. Moreover, high-fidelity weak measurements would require quantum efficiency higher than what has been obtained in previous demonstrations of weak continuous measurement \cite{murch2013observing,weber2014mapping,minev2019catch}. Preliminary experiments of the direct parity measurements that are needed here have been attempted \cite{riste2013deterministic,roch2014observation}, but have not yet achieved the fidelity required for a convincing demonstration of the quantum pigeonhole effect. We consider this proposal as a challenge to the experimental community in the near term.

\section{The past of quantum particles}

The failure of the pigeonhole principle for quantum particles happens when they are pre- and postselected, so the failure takes place in the past. However, standard quantum theory does not provide a clear picture for the past of quantum particles. In our discussions of the quantum pigeonhole principle, we adopted a definition of a counterfactual character: if it was inferred using the ABL formula (\ref{ABL}) that the particle would be found in a particular box with certainty when searched, then we said that the particle was in the box. This definition has a conceptual difficulty that we must address: we want to make claims about the presence or absence of quantum particles in a box even when we do not check the box. One could argue that a strong measurement, in fact, changes the quantum state and thus disrupts the ability to make the desired inference.

Vaidman \cite{past,Universal} proposed an alternative way to reason about the past of a quantum particle that can be grounded more operationally: if the particle was there, it should weakly interact with the environment and leave a trace behind. Thus, the question of a particle's presence in a box can be answered by checking for the presence of a trace left in the box after the postselection. This will not be a robust trace in which the environment changes its state to an orthogonal state, since that would disrupt the evolution of the particle too sharply. The appearance of a small amplitude of an orthogonal component for the environment is sufficient to establish a suitable trace. More precisely, the amplitude of the orthogonal component in the environment should be of the same order as the amplitude of the component that would appear if a single well-localized particle were placed in the box directly.

To simplify the analysis, we assume that for each particle there is a particular (different) position at which it could reside in the box, so any trace it could leave in the box will be independent of traces from other particles.
Every particle that is definitely in the box leaves a trace with a small amplitude $\epsilon$, i.e. the local environment of a particle $j$ makes the following evolution:
\begin{align}\label{env}
|\chi_{j} \rangle \rightarrow  \eta\, | \chi_{j} \rangle + \epsilon\, |\chi_{j} ^{\perp} \rangle .
\end{align}
We assume that nothing happens to    the environment if there is no particle.

With these assumptions for the trace, all discussed setups for the quantum pigeonhole paradox will show a trace of order $\epsilon$ for every particle in every box after the postselection. That is, there will be amplitudes of order $\epsilon$ for every orthogonal component state $|\chi^{\perp}_j \rangle$. Thus, according to Vaidman's weak trace criterion, every particle indeed was present in every box. This is already a somewhat paradoxical situation, but it is not a demonstration of the failure of the pigeonhole principle. As is well-known from Hardy's paradox \cite{Hardy,V-HardyPRL}, we cannot conclude that the two particles are present together in a box solely from the evidence of single-particle traces.
%
%
%
If we put a pair of particles $j$ and $k$ into box $A$, the interaction with the environment described by (\ref{env}) leads to the appearance of a component with the product state $|\chi^{\perp}_j \rangle|\chi^{\perp}_k \rangle$ with an amplitude of order $\epsilon^2$. This is the proper criterion for particles $j$ and $k$ both occupying a box together.

Let us analyze our examples according to this criterion. Are particles 1 and 2 present in box $A$ after pre- and postselection in the example of four particles in two boxes, with Eqs. (\ref{eq:1}, \ref{eq:2})? The component of the state of the environment in box $A$, $\epsilon^2|\chi^{\perp}_1 \rangle_A|\chi^{\perp}_2 \rangle_A$, is created due to the interaction, but it disappears after the postselection. However, even after the postselection, we are still left with a trace of order $\epsilon^2$ in box $A$ of particles 1 and 3 and, in fact, of any other pair except for 1 and 2. Thus, according to the naive two-particle trace criterion, there is no failure of the pigeonhole principle after checking all pairs in this example. This fact helps clarify the limited meaning of this example of the paradox: we have to consider only a special measurement interaction in which the environment does not distinguish between situations in which different nonzero numbers of particles are present in the box. Only in that case will all the traces cancel to correspond to the conclusion from the ABL rule (\ref{ABL}).

In contrast, direct local traces demonstrate the failure of the pigeonhole principle of example \cite{peculiar2013} well. If we arrange a situation in which only two particular  particles, $j$ and $k,$ leave local traces in the boxes according to (\ref{env}), then after the postselection there will be zero amplitude for the two-particle trace $\epsilon^2|\chi^{\perp}_j \rangle_A|\chi^{\perp}_k \rangle_A$ and zero amplitude for the two-particle trace $\epsilon^2|\chi^{\perp}_j \rangle_B|\chi^{\perp}_k \rangle_B$. In a realistic case, when all particles leave amplitude $\epsilon$ traces, the two-particle trace will survive the postselection, however only as a part of three-particle trace, such as  $\epsilon^3|\chi^{\perp}_j \rangle_A|\chi^{\perp}_k \rangle_A |\chi^{\perp}_l \rangle_A$. Since the amplitude is of the third order in $\epsilon$ (and not the second), such a trace is neglected.

In the example \cite{aharonov2016},  similarly to our four-particle example, local two-particle traces $\epsilon^2|\chi^{\perp}_j \rangle_A|\chi^{\perp)}_k \rangle_A$  persist for every pair of particles in box $A$ (with similar traces in box $B$). This example thus represents a failure of the pigeonhole principle with a limited meaning. We can claim that the particles in every pair do not occupy the {\it same} box, but we can only check this statement with a measurement that does not tell us which specific box. The measurement thus cannot be just some local measurement; it requires entanglement of the environment in box $A$ and box $B$ such that the local couplings leave the box identities uncertain. To this end a nonlocal parity measurement is needed for a proper demonstration.

The measurement should arrange local couplings in box $A$ and $B$ such that particle $j$ in box $A$ affects a composite entangled system I of local environments in $A$ and $B$ exactly in the same way as particle $k$ in box $B$, and also that particle $j$ in box $B$ affects another composite entangled system II of the environment exactly in the same way as particle $k$ in box $A$.
In this situation,  particles $j$ and $k$ in box $A$ will create trace  in the environment $\epsilon^2|\chi^{\perp}_j \rangle_{\rm I}|\chi^{\perp}_k \rangle_{\rm II}$, but this trace will be exactly the same as the trace left by the two particles present in box $B$, and thus the environment will know that the two particles are in the same box, but will not know in which. For such a specially tailored environment there will be no trace  $\epsilon^2|\chi^{\perp}_j \rangle_{\rm I}|\chi^{\perp}_k \rangle_{\rm II}$ after the postselection, in correspondence to the claim that there are no particles in the {\it same} box. Parity measurement procedures of this type were described in the previous section.

The trace approach to the past of the particle allows us to understand the failure of the pigeonhole principle for pre- and postselected quantum systems from an operational perspective, but it requires careful reasoning. For quantum systems we cannot apply classical arguments according to which if both particles $j$ and $k$ are each in box $A$, then they are both together in $A$. The operational meaning of ``the particle was in the box'' is that the particle left a single-particle trace in the box. Similarly, the meaning of ``two particles were in the box together'' is that there is a two-particle trace left in the box (as opposed to two single-particle traces). When the particles are pre- and postselected, there are cases with two single-particle traces in the box without a corresponding two-particle trace in the box. (Note that this possibility requires some entanglement between the two single-particle traces.) These cases correspond to the failure of the pigeonhole principle and is in agreement with the original approach according to which we decide that a particle is in the box if it could be found there with certainty, and two particles in the box if they could jointly be found there together with certainty.

\section{Conclusions}
We have carefully revisited in what sense the classical pigeonhole principle may be violated by quantum systems. We introduced a variation of the traditional quantum pigeonhole paradox that corresponds more directly to the statement of the original pigeonhole principle. However, our variation will be challenging to implement experimentally due to the need for threshold-type detection that is largely insensitive to pigeon number. The existing  examples, that show failures of consequences of the pigeonhole principle instead, have the advantage of being easier to implement experimentally.

Although several experiments exist already for these simpler examples, our careful examinations reveal several shortcomings that prevent them from being definitive demonstrations of the paradox. To address these shortcomings, we suggested alternative experimental implementations that would more convincingly demonstrate the failure of the pigeonhole principle. In particular, using direct parity measurements of superconducting transmons seems like a promising way to compellingly demonstrate the paradox with mesoscopic quantum hardware at the near-classical micron scale.

The quantum pigeonhole paradox traditionally uses the ABL rule to establish elements of reality in the past.  We argue that these inferences correspond perfectly to a more empirically grounded test, namely the identification of weak environmental traces. That is, pigeons will leave behind footprints in the boxes that one can later detect. Since such weak traces are detectable in experiments, this criteria gives a firm operational meaning for the paradox.

\begin{acknowledgments}
  This work has been supported in part by the  U.S.-Israel Binational Science Foundation Grant No. 735/18, and by PBC PostDoctoral Fellowship at Tel Aviv University.
\end{acknowledgments}

%
\end{document}